# Intrinsic properties of suspended MoS2 on SiO2/Si pillar arrays for nanomechanics and optics.

Julien Chaste[1*], Amine Missaoui[1], Si Huang[1], Hugo Henck[1], Zeineb Ben Aziza[1], Laurence Ferlazzo[1], Carl Naylor[2], Adrian Balan[2], Alan. T. Charlie Johnson Jr.[2], Rémy Braive[1,3], Abdelkarim Ouerghi[1]

[1] Centre de Nanosciences et de Nanotechnologies, CNRS, Univ. Paris-Sud, Universite Paris-Saclay, C2N – Marcoussis

[2]Department of Physics and Astronomy, University of Pennsylvania, 209S 33rd Street, Philadelphia, Pennsylvania 19104 6396, United States

[3]Université Paris Diderot, Sorbonne Paris Cité, 75207 Paris Cedex 13, France

* Corresponding author: julien.chaste@c2n.upsaclay.fr

**Abstract**

Semiconducting 2D materials, such as transition metal dichalcogenides (TMDs), are emerging in nanomechanics, optoelectronics, and thermal transport. In each of these fields, perfect control over 2D material properties including strain, doping, and heating is necessary, especially on the nanoscale. Here, we study clean devices consisting of membranes of single-layer $MoS_2$ suspended on pillar arrays. Using Raman and photoluminescence spectroscopy, we have been able to extract, separate and simulate the different contributions on the nanoscale and to correlate these to the pillar array design. This control has been used to design a periodic $MoS_2$ mechanical membrane with a high reproducibility and to perform optomechanical measurements on arrays of similar resonators with a high-quality factor of 600 at ambient temperature, hence opening the way to multi-resonator applications with 2D materials. At the same time, this study constitutes a reference for the future development of well-controlled optical emissions within 2D materials on periodic arrays with reproducible behavior. We measured a strong reduction of the $MoS_2$ band-gap induced by the strain generated from the pillars. A transition from direct to indirect band gap was observed in isolated tent structures made of $MoS_2$ and pinched by a pillar. In fully suspended devices, simulations were performed allowing both the extraction of the thermal conductance and doping of the layer. Using the correlation between the influences of strain and doping on the $MoS_2$ Raman spectrum, we have developed a simple, elegant method to extract the local strain in suspended and non-suspended parts of a membrane. This opens the way to experimenting with tunable coupling between light emission and vibration.

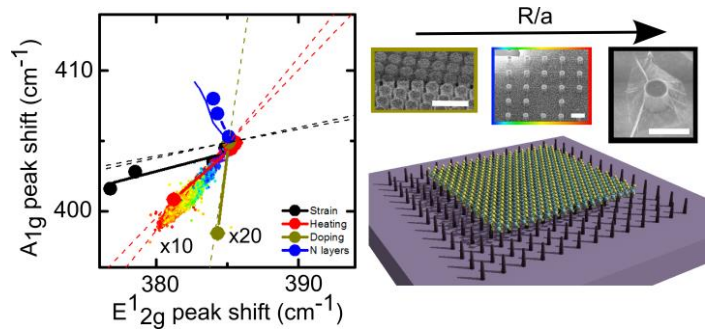

Bi-dimensional (2D) materials (such as graphene, transition metal dichalcogenides (TMDs) *e.g.* $MoS_2$ and $WSe_2$) are representative of 'flatland' technology. Due to the 2D nature of these materials, their intrinsic properties are greatly influenced by the surrounding environment on the nanoscale. Within these systems, the suspended structures [1–3] are hybrid systems with transport performance, [4–9] tunable optoelectronics behavior [10–14] and nanomechanical properties [15–17] that are all interesting on their own, but which can be coupled together, for example, the coupling of a vibration with a localized quantum emitter. Only a few systems at the interface between nanomechanics and nano-optics have achieved a similar richness of possibilities. [18,19] Control over strain, doping, and temperature in 2D materials is vital for both the nanomechanical performance and the localized optical emission. Achieving such control is directly related to sample quality and to local nanostructuring of the suspended membranes.

Up to now, the experimental realization of suspended nanomechanical resonators has focused mainly on multilayer $MoS_2$ [15–17,20–22] without specific nanostructuring and with single resonator systems. Nevertheless, in order to achieve the next step in the engineering of 2D structures, it is necessary to combine high-quality 2D membranes with a controlled nanostructuring process.

In parallel, localized quantum dots and efficient quantum emitters made by strain nano-engineering of TMDs [11–14] have been measured using peak arrays on a substrate for non-suspended 2D materials. TMDs are suitable materials for such optical applications because of their semiconducting properties with different band gap energies and the possibility to finely tune their doping. They can also be easily combined to form interesting heterostructures with various properties. When a TMD membrane is distorted by nanopillars, the strain will strongly influence the band structure and hence allow the creation of confined modes for electronics and optics (typically, 1% of strain reduces the band gap by about 0.1 eV).[23,24] Specifically, in suspended samples, static strain and band structure can be tuned by applying an electrostatic potential using an embedded back gate. Such a design, consisting of a membrane suspended on nanopillars, represents a platform for the creation of highly efficient and tunable optoelectronic devices. Nonetheless, a good understanding and control of the strain, doping, and temperature on the nanoscale are essential. In particular, it is fundamental to understand the mechanism of the direct to indirect bandgap transition in monolayer TMDs under strain. Here we use suspended membranes on nanopillar arrays to elucidate these aspects. Although $MoS_2$ is subjected to a small broadening of its emission peaks, it has distinct Raman features in contrast to $WSe_2$. Moreover, $MoS_2$ has greater resistance to oxidation and is a referenced material among the TMDs, especially for Raman spectroscopy, nanomechanics, nanostructuring by nanopillars and photovoltaic [23,24] or photocurrent generation.[25]

Here, we study a system composed of suspended MoS$_2$ membranes deposited on nanopillar arrays in order to find a correlation between nanostructure dimension and strain, doping and temperature variations. We use a device with a monolayer deposited directly on top of SiO$_2$/Si nanopillar arrays acting as a "flying carpet", where the MoS$_2$ layer itself consists of suspended and non-suspended parts.

**Results**

Thanks to chemical vapor deposition, very large 2D membranes (>100µm) of monolayer MoS$_2$ were used. We systematically studied the effect of pillar geometry on the strain, doping, and heating of the MoS$_2$ flakes by Raman spectroscopy and photoluminescence. We noted that, in some parts of the sample, periodic ripples, linked to pillar position, had been created in the MoS$_2$ flakes. The effect of these ripples was correlated to a strain effect in the Raman spectra.

In order to strongly enhance this strain effect, we spatially separated the pillars. In this configuration, it was possible to obtain a "tent" structure around one or more nanopillars. The 2D membrane sat on the tops of the pillars, thus creating a high strain gradient along the tent. This makes the structures suitable for obtaining a strong piezo-phototronic effect in 2D materials and gives rise to artificial atomic states within the 2D membranes.[24] We have also observed with Raman spectroscopy, the appearance of indirect band gaps induced by high strain and nanostructuring in monolayer MoS$_2$.We managed to optically separate strain, heating and doping effects using Raman spectroscopy and to control these by varying the device geometry itself. Clear evidence of a correlation was highlighted between the observed behavior and the aspect ratio R/a (where R is the peak radius and a, the peak separation length).

A suspended monolayer of MoS$_2$ (triangle shaped) over a large array of SiO$_2$ pillars is shown in the diagram in Figure 1. The sample was made using a method similar to Reserbat-Plantey *et al*.,[26] with doped Si/SiO$_2$ as substrate. The pillar mask was made using e-beam lithography, followed by Ni deposition. The pillars were formed using dry etching of the SiO$_2$/SI substrate before MoS$_2$ deposition and the Ni mask was completely removed afterward by wet etching. Two different samples were measured, labeled A and B, respectively, with SiO$_2$ thicknesses of 1450 nm and 670 nm, respectively. The pillar heights were 820 nm for sample A and 420 nm for sample B. In sample B, the optical reflection at 532 nm of the MoS$_2$-substrate cavity was optimized (see SI).[27] Large monodomains of MoS$_2$ were grown by chemical vapor deposition (CVD) on a non-patterned Si/SiO$_2$ substrate and then transferred to the pillar arrays by wet transfer, as described in previous studies,[28,29] After a few cycles of cleaning in water, the MoS$_2$-PMMA was transferred on top of the pillar substrate, the resist dissolved in acetone and the sample dried in a critical point dryer. The results with different R/a ratios are presented in Figures 1 and SI. Using this technique, we managed to obtain large membranes (about 6.5 µm long) of fully suspended MoS$_2$, as shown in Figure 1c.

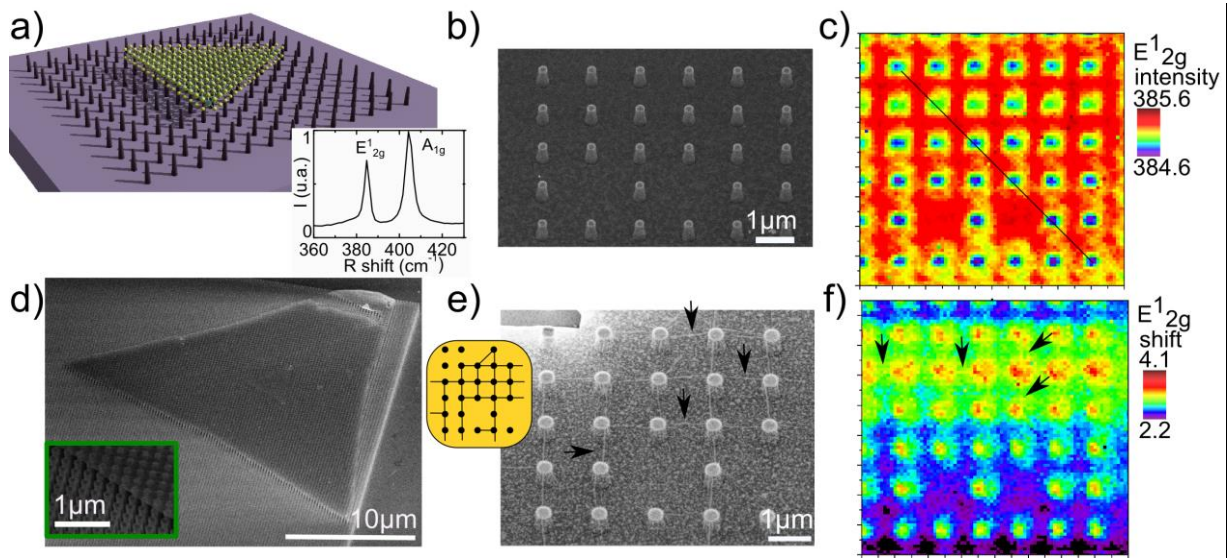

**Figure 1**: **MoS$_2$ on nanopillar arrays:** a) Diagram of a suspended MoS$_2$ layer deposited over a SiO$_2$ pillar array and inset showing a typical Raman spectrum with the E$^1_{2g}$ and A$_{1g}$ peaks in the suspended region. b),d) and e) E-beam image of typical flakes of MoS$_2$ fully suspended over pillar arrays. In e), lines and ripples at specific positions correlate with the pillar array pattern, as also seen in the inset for clarity. In this area, the ripple number is close to covering 100% of the surface. In c) and f), Raman mapping of the E$^1_{2g}$ peak position shift and intensity along the SiO$_2$ peak array are shown. Periodic patterns are seen in correlation with the SiO$_2$ peak position. The arrows in f) highlight some lines in the peak position of E$^1_{2g}$ mapping relying on the different pillars and attributed to similar ripples and strain patterns as in e). These lines are not visible in the peak position of A$_{1g}$.

In Figure 1, organized ripples can be seen clearly connecting the peak apices in the suspended membrane that appear in some regions of the samples. Completely planar features were also obtained. Generally, the organized ripples appeared preferentially at the borders of the MoS$_2$ flakes, while the central areas stayed relatively flat. These ripples were also observed in the case of non-suspended graphene deposited on similar pillar surfaces.[26] These ripples originated from the stress created at the peak apex and distributed along the membrane.[26] In fact, the model of these mechanical instabilities can be described by a suspended Föppl-Hencky membrane with a clamped boundary condition and submitted to a point load. [24,30] This is a good indication that strain control on the nanoscale is possible with peak array engineering.

Hereafter, the Raman measurements obtained at ambient conditions on SiO$_2$ pillar arrays fully covered by a very clean and uniformly suspended monolayer MoS$_2$ membrane are presented and discussed. The main apparent feature is correlated to pillar patterns in both peak intensity and position: there are physical contact and interaction between the MoS$_2$ and the top of the pillars. We identified a strain signature along the ripple positions in the MoS$_2$ Raman maps of the E$^1_{2g}$ mode (Figure 1e). This effect was absent in the A$_{1g}$ peak position. In order to determine the strain at the pillar apex, where it must be of maximum amplitude, the contributions from doping, heating, and strain had to be separated. To do so, one identified the Raman dependency of each contribution; here we propose another possibility by engineering high strain in a tent-like structure.

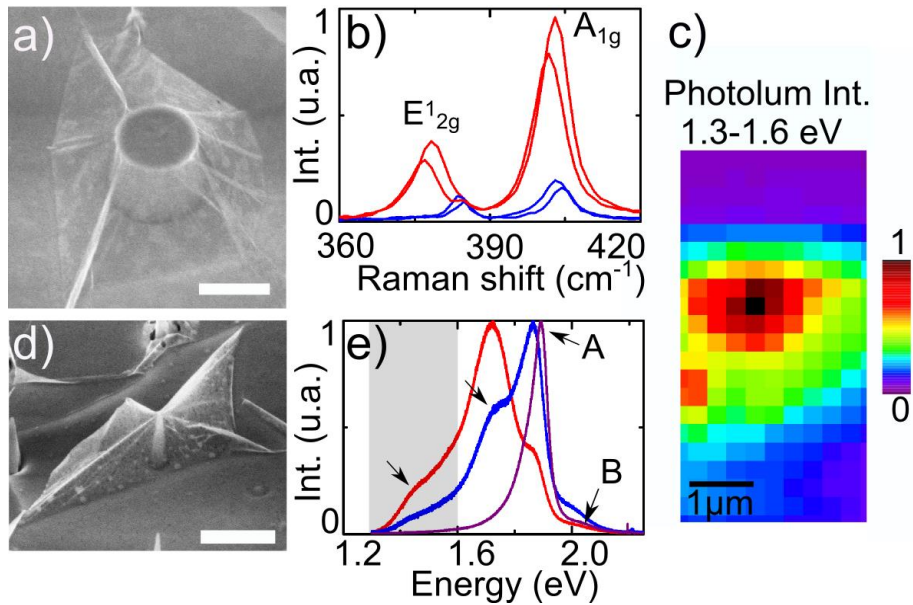

**Figure 2: Strain engineering with C2N samples:** a) and b) A "tent" structure made of CVD MoS$_2$ deposited on SiO$_2$ pillars in the limit where 2R< a. b) Raman measurements for highly stressed MoS$_2$, pinched by SiO$_2$ pillars. A shift of 8 cm$^{-1}$ is observed for the E$^1_{2g}$ between the reference and the peak position at the pillar apex and represents a minimum planar stress of 2%. We assume the difference in intensity is mainly due to the suspended/not-suspended MoS$_2$. c), Photoluminescence signature of the MoS$_2$ structure, shown in d) with the signal integrated between 1.3 and 1.6 eV. We observe reduced band gap energy in the strain region, around the pillars. e), some extracted spectrum at the peak apex (red), at a closed position (blue), and at a reference point (purple) (all normalized). At the peak apex, the strong shift of the A and B peaks corresponds to high band gap reduction and a minimum of 2% elongation. We see the appearance of another peak at lower energy, indicated by the black arrow, which we attribute to the indirect band gap in the MoS$_2$ monolayer under high strain.

Figure 2 shows MoS$_2$ flakes, which are pinched by a single SiO$_2$ pillar. This pillar is three times higher than that obtained in similar studies, allowing the strain limits to be reached in such devices. MoS$_2$ suspended over 2 or three pillars in these tent structures are also presented in Supplementary Information. Using this pillar height, crack propagation in the MoS$_2$ was observed on the surface. This indicated that the limit of strain in MoS$_2$ had been reached. In Figure 2b, we present two extreme cases of strain signature at the pillar apex, using Raman spectra. The E$^1_{2g}$ peak has shifted markedly, by almost 8 cm$^{-1}$ on the pillars. Using the proportionality $\Delta A_{1g}= 0.26.\Delta E^1_{2g}$ (see SI), we calculated a strain of about 2%. Doping and temperature can be excluded, as these cannot be linked to this behavior. This strain estimation is undervalued due to the optical spot size resolution, around 300-400 nm, and the effective broadening of the resulting signal. We assumed that the maximum strain must be reached locally at the nanopillar apex (tens of nanometers). In Figure 2e, the photoluminescence response was measured along a sharp tent structure. Three resonances are visible: A$^-$, A and B (around 1.82, 1.85 and 2.0 eV, respectively). A and B correspond to the spin-orbit coupling split of the excitonic resonance and A$^-$ corresponds to the MoS$_2$ trion binding energy. Around the pillars, we observed a global resonance shift and large broadening with complex features. This corresponds to a strong band gap reduction of the MoS$_2$ monolayer, as previously observed. [31] In our case, the main A peak position, shifted to 1.72 eV. It corresponds to a strain increase of 2.4% around the main structure (see **SI**). No strong evidence of any Raman peak deviation was observed at the same position. We explain this difference using the so-called phenomenon of exciton funneling. [23,24] In contrast to Raman processes, photoluminescence is directly related to optical excitons and can be subjected to diffusive effects which indirectly enhance the emitted signal of the MoS$_2$ strain. This appears mainly at ambient temperature, where the created excitons diffuse to the pillar apex position, with the minimum band gap value, before emitting a photon at this lower energy. Broadening of the peak indicates it is not fully effective over the length of the spot size and we obviously see different contributions. The optical reflectance is modulated by the MoS$_2$-substrate cavity (height $d_{air}$). At less than 0.5μm around the "tent" structure, $d_{air}$ varies from 600-800 nm to 0 nm and this cavity effect is completely smoothed after convolution with the laser spot. Also, the cavity quality is not enough to enhance the signal dramatically and is not related to the emergence of a strong signal between 1.3 and 1.6 eV in Figure 2.

Closer to the pillar position, an additional contribution appeared at a lower energy and shifted down to about 1.45 eV with an increased shift variation, when compared to the main intensity position. We attribute this peak to the indirect band gap contribution appearing in monolayer MoS$_2$ with no biaxial strain above 1.5-2 %.[33–43,25] This had been observed previously on the nanoscale[32] on a rough substrate with a complex strain tensor and on a smooth substrate, as reference.[31] In this study, the authors did not notice any additional contributions related to an indirect band gap signature in a suspended MoS$_2$ monolayer under ultra-large biaxial strain of up to 5.6%. Moreover, the indirect band gap resonance was expected to be less luminescent than a direct process. Compared to the data presented here or ref.,[32] our situation was quite different as the nanostructuring gave rise to a strong strain gradient and involved fewer processes, which increased the indirect gap response: 1) The consequences of the exciton funneling effect are not easily determined in this configuration and can favor indirect band gap processes. 2) Suspended MoS$_2$ has a more intense photoluminescence signal than non-suspended and our suspended part was the stressed region. This improved the indirect peak intensity in our case. 3) Finally, there must be an intermediate regime between uniaxial with random orientation [35] and biaxial strain, which results also in divergence from ref.[31] and affects the modulation of the peak position or intensity *versus* strain.

**Discussion**

In order to separate, quantify and simulate the 2D properties of a typical Raman pattern, as in Figure 1, it was necessary to understand their contributions to the Raman peak features. For this, different situations were studied where strain, doping, temperature and the number of layers were clearly identified and separated from the other contributions. We focused on the flake of Figure 1c (sample B) with different contributions;

1) The signal position for both the E$^1_{2g}$ and A$_{1g}$ peaks are shifted by almost 0.4 cm$^{-1}$. In Figure 3c, a strong linear dependence is seen between the position of the two peaks E$^1_{2g}$ and A$_{1g}$ with a slope of about 1.1. This indicates a thermal heating effect and should lead to a thermal transport simulation.

2) The observed A$_{1g}$ peak width and peak intensity modulation, which is un-correlated to the E$^1_{2g}$ peak features, suggested a doping variation. It is reasonable to assume a doping modulation along the sample due to substrate interaction. Doping contributions on the Raman spectrum appear mainly at the A$_{1g}$ peak position.[43] We defined $n_n$ as the doping of the MoS$_2$ layer on the pillars (non-suspended) and $n_s$ elsewhere (suspended). We determined empirically that the main dependence in our results came from the difference $\Delta n = n_s - n_n$.

3) In parallel, some lines are visible between the pillars in the peak position of $E^1_{2g}$, which are related to organized ripples and strain; this is less visible in the $A_{1g}$ mapping. Strain at the pillar apex in our sample must not be neglected but can still be avoided if only the monolayer MoS$_2$ $A_{1g}$ peak dependence is considered. Once the other parameters are well known, the stress at the apex can thus be estimated.

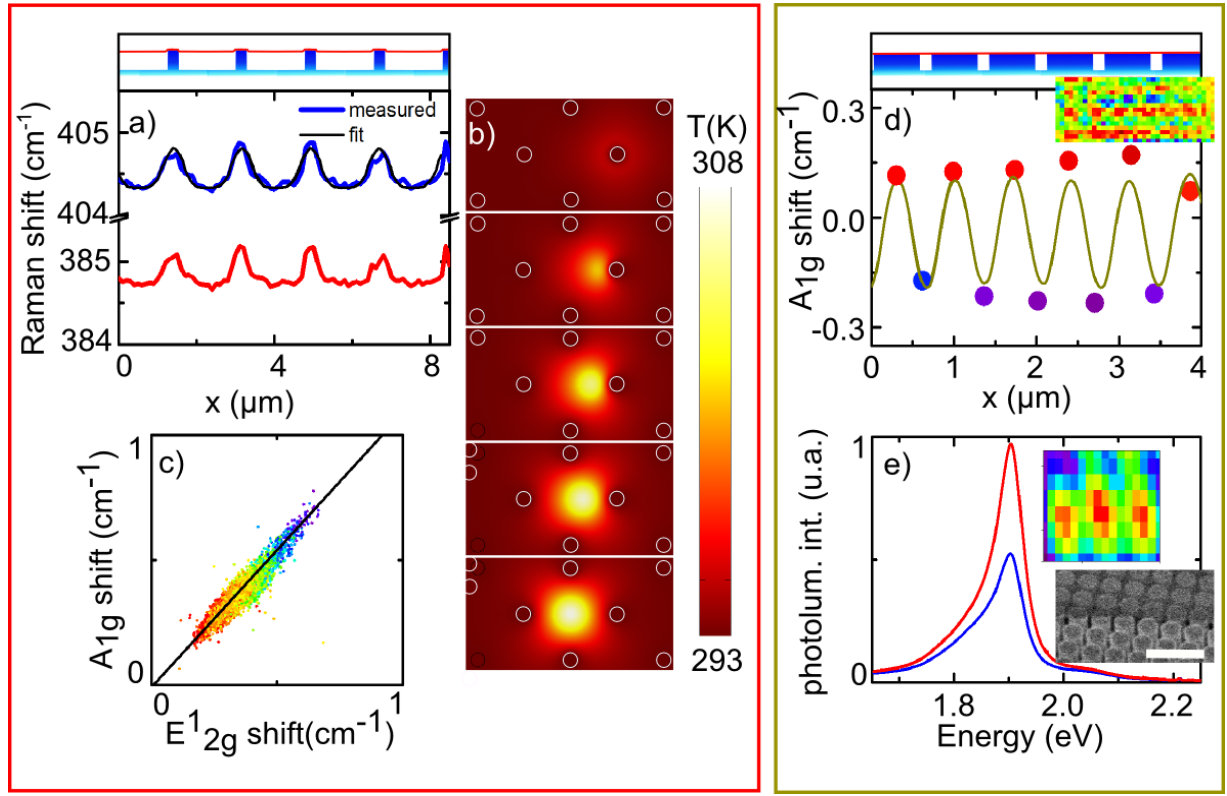

**Figure 3: Clean periodic patterns in Raman signatures of the MoS$_2$** a) Fast Raman mapping of the $E^1_{2g}$ and $A_{2g}$ peak position shift and intensity along the SiO$_2$ peak array along the black line in Figure 1c, without any correction. A fit of the $A_{2g}$ peak position is also plotted, taking into account the doping difference of MoS$_2$ due to substrate interaction and thermal heating, with a thermal conductivity of 50W.m$^{-1}$.K$^{-1}$. c) A Lee diagram, [45] adapted to MoS$_2$, for the points in Figure 1c, with the corresponding color mapping (For clarity small deviations have been removed due to our 15h of measurements). The slope, close to 1, scales well with a temperature dependent signature. The broadening of this plot in the middle part and the curved shape in b indicate, in fact, a more complex schematic with a melting of strain, doping, and temperature. **Doping modulation** with a sample 2R~a. In d), Raman $A_{1g}$ peak modulation along the sample with, in the inset, the corresponding mapping. This modulation indicates a doping change of about $5.10^{12}$ cm$^{-2}$ with a spot size of 300nm. After correlation with the SiO$_2$ peak intensity we deduced the doping in the suspended part to be higher than in the non-suspended part. e) Photoluminescence spectrum for two positions, on top of a peak (non-suspended) and between the peaks (suspended) with, in the inset, an image of the sample and mapping of the maximum intensity. Scale bar is 1µm. Only the intensity of the A peak is modulated.

In order to determine the doping variation Δn, we studied a different situation with large 2R≈a, where doping was clearly identified and separated from other contributions (Figure 3d). Surfaces with suspended and non-suspended MoS$_2$ were equivalent and the MoS$_2$ seemed flat. On the raw data (see **SI**), a modulation of the $A_{1g}$ peak position can be observed along the sample, which is less effective in the $E^1_{2g}$ position. As the signal is quite small, we have to carefully remove the background deviation and average our periodic data to extract the exact $\Delta E^1_{2g}$ and $\Delta A_{1g}$ dependence. We obtained $\Delta A_{1g} = 8.2\, \Delta E^1_{2g}$, which is definitively a doping dependence signature. In the SI, we proposed a calibration of this value on a non-suspended sample. One limitation of the optical method is the diffraction limit, which has almost the same length as our nanostructures. To avoid this limitation and to be more quantitative, we used an innovative procedure for data analysis. Due to the finite spot size, we took into account that the local doping distribution and the resulting local spectral Raman response were spectrally convoluted with the Gaussian distribution of the laser and the collected photons. The laser diameter was determined with *in-situ* calibration during the same measurement at the $MoS_2$ flake edge, equivalent to 300 nm. This gave this sample a higher doping score in the suspended part; $\Delta n = +5.10^{12}\,cm^{-1}$. This high doping was also confirmed by the very strong photo-gating observed (see SI) and by the shape of our $MoS_2$ triangles, which suggested sulfur vacancies.[44]

For the sample in Figure 1c, once the doping on the $MoS_2$ was known, it was possible to extract the thermal conductivity from the analysis of the $A_{1g}$ position mapping, because the strain effect is negligible on this peak. For the temperature contribution, the local heating, at one point, was obviously induced by the laser itself, and the related Gaussian distribution of the absorbed power. This means that the temperature distribution along the sample was different for every measurement point and it was modeled by finite element analysis with COMSOL software, as in Figure 3b. We naively considered here a thermal transport to be within the Fourier law limit. It is a simple but good approximation for thermal transport in 2D membranes [4] if we stay in the low heating regime and the typical length scale, between the heat source and the thermal bath, is more or less constant in the measurements presented here. We consider the pillars to be in a perfect thermal contact with the $MoS_2$, otherwise, no temperature gradient along the $MoS_2$ would be present nor any effect at all in the Raman data, which was not the case. Our data fit quantitatively and qualitatively well with a spot size of 400 nm and a thermal conductivity of 50 $W.K^{-1}.m^{-1}$, as shown in Figure 3a. The laser spot size is in agreement with previous results. In the SI, we have calibrated the $\Delta A_{1g}/\Delta E^1_{2g}$ ratio for temperature variation on a non-suspended sample. We used a doping difference of $\Delta n = +5.10^{12}\,cm^{-1}$ and a laser power of 50μW and estimated the heating to be at a maximum of around 15 K at when the laser was far away from the pillars. The planar thermal conductivity corresponded to typical values in the literature for suspended monolayer $MoS_2$ and confirmed our methodology to be appropriate. It must be noted that different $\Delta n$ were also used to fit our results, with less success. This directly confirms the value of $\Delta n = +5.10^{12}\,cm^{-1}$ to be general in our samples.

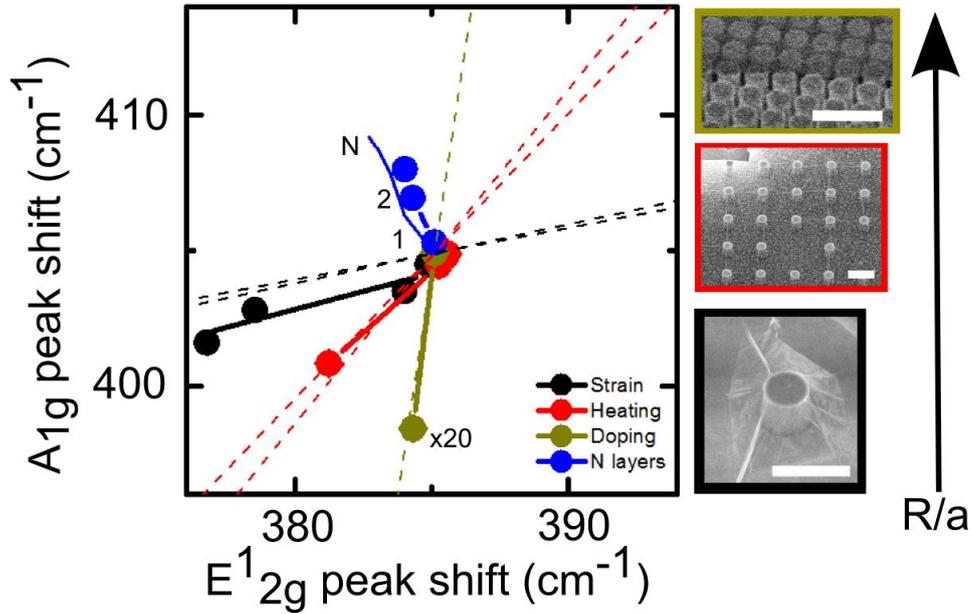

**Figure 4: A Lee et al diagram [45] for our different sample configurations for strain, doping temperature and layer number.** A plot of the $A_{1g}$ peak positions in function of the $E^1_{2g}$ peak position is presented for four different cases described in this paper. Blue circles show the number of layer variations compared with the blue reference line (data extracted from).[49] Black circles show strain variation around the tent structures of Figure 2 compared to the expected slope from the literature (solid black line) and the dashed line for the strain. Red circles show the measurement of heating with the laser itself and brown circles, the doping induced by the pillars.

For this measurement, temperature contributes only about 33% of the total $A_{1g}$ shift position and doping, the rest. With this, the strain at the pillar apex can be calculated from the Raman measurements. For this, we used an analogy between the $A_{1g}$ doping variation and the $E^1_{2g}$ strain variation. In Figure 4, shown using a Lee et al. diagram,[45] the different cases, with strain, temperature, doping and layer number, with the respective slopes expected from the literature in a dashed line (see SI). Doping and strain variations were extracted from samples and data in Figures 2 and 3, respectively. Temperature variations were obtained for a suspended $MoS_2$ similar to that in Figure 1c, far away from the $SiO_2$ pillars, for different laser strengths (10 µW to 500 µW). We can estimate the $\Delta A1_g / \Delta E^1_{2g}$ to be around 1.1 as expected in the literature, for thermal heating of a $MoS_2$ membrane (see SI for references and our own calibration of heating and doping). Three observations can be drawn from these results. First, we confirm that the measurements were done on monolayer $MoS_2$ for all the data presented, as they have same origins for all the measurements, in accordance with the monolayer $MoS_2$ case (blue point). Second, the doping, strain, and temperature variations can be optically separated in $MoS_2$ without any ambiguity. Finally, we have a quasi-mirror symmetry situation between the $A_{1g}$ and the $E^1_{2g}$ for strain and doping effects, respectively, to the slope of 1. This diagonal corresponds, coincidentally, to the temperature variation.

From this and the slope of about 1 obtained in Figure 3c, it is possible to estimate the strain contribution of about 66% of the peak shift. This mixing of the various contributions with different Raman variations explains, in the Lee et al. diagram [45] of Figure 3c, the broadening of the data in the center region as the strain and doping variations are not completely alike and without the exact same spatial distribution. Nevertheless, it seems that strain and doping contributions are equivalent in a first approximation. In the case of strain, it is necessary, first, to simulate the strain distribution along the sample with a model that is not trivial in such a suspended configuration. [24] Taking advantage of this analogy between strain and doping description, the strain description can be simplified to a simple strain difference $\Delta\varepsilon$ between the suspended $\varepsilon_s$ and non-suspended region $\varepsilon_n$, where $\varepsilon_s$ and $\varepsilon_n$ are assumed to be constant in these regions. With this assumption, there is a real local difference $\Delta\varepsilon$ of around 0.5% between the pillar apex and the suspended area of the sample. It should be noted that in this situation, where strain and doping are symmetric and similar, this configuration can now be used subsequently and extended to more complex features in order to understand more general patterns.

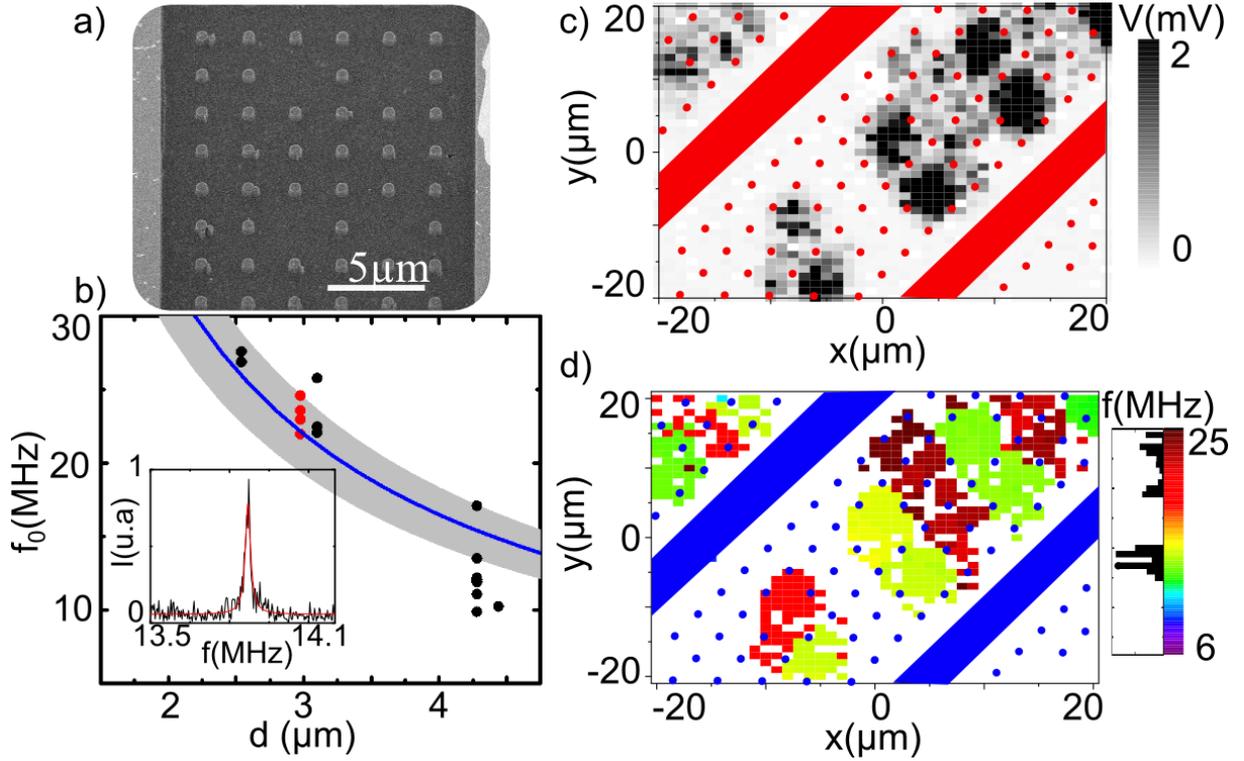

**Figure 5: Optomechanical measurement of an array of MoS$_2$ resonators.** a) An E-beam image of suspended MoS$_2$ over a pillar array with electrical gold contacts S and D on both sides. We used the substrate as a gate G to electrostatically excite the vibration. b) Resonant frequencies as a function of the membrane diameter locally defined by the pillars. Inset shows an example of a resonance with Q~600 even at ambient temperature. The blue line corresponds to a strain of 0.045% ±0.01% on the suspended part of the MoS$_2$. c) and d) spatial mapping of mechanical resonance amplitude and frequency, respectively ($V_G$=5V, $V_{G,ac}$=10mV, $V_{DS}$=0V, He-laser spot size=4μm, laser power =12μW ).

In Figure 5, we show optomechanical measurements using a quasi-similar set-up, as described in reference .[15] Here, MoS$_2$ flakes were suspended over pillar arrays, with electrical contacts on each side. This pillar array configuration was previously proposed by Midtvedt *et al.* [46] in order to create non-linear phononic elements within 2D materials. Each individual membrane was formed locally with 4 pillars or more at the corners. With our samples, we were able to excite electrostatically the vibration with a back gate and to measure it optically with the reflected signal, at 633 nm, a Michelson interferometer and a photodetector. In the inset of Figure 5c, we observe a mechanical resonance of around 13 MHz with a record quality factor ~600 for a monolayer MoS$_2$ vibration at ambient temperature. In Figures 5b and d, we present the amplitude and frequency $f_0$ of the resonance extracted from a spatial mapping of our resonator array.

Even with our large spot size~4μm, we clearly observe a periodic resonator in between the pillar positions. Moreover, the mechanical frequencies are close to each other if we consider similar membrane shapes and dimensions, with low-frequency dispersion. If we analyze the resonances $f_0$ for different membrane diameters d in different flakes and samples, we find a good match with the Euler Bernoulli description of motion [17] and with a stress of 0.045% in most of our resonators. This measurement is another factor which highlights the high quality of our devices. A 2D series of resonators has been measured in TMDs and it is proof of the concept that this 2D material can be used for phononic applications only when strain, doping, and heating are well controlled in a periodic device. The mean strain $ε_S$ of 0.045% on the suspended part is a measure for a membrane of dimension $d_S$ (~3μm). With the following simple argument, we can deduce the strain on the pillars of diameter $d_N$ (~280nm). If we consider the elongation between the suspended and the non-suspended parts to be more or less equal, we have $ε_N$~ $ε_S.d_S/d_N$=0.48% which is close to the measured value of 0.5% with Raman. In fact, for a fixed $ε_N$ at the pillar position, we can estimate the membrane strain to diminish when we increase $d_S$. This is exactly what we observed: for long membranes, the static strain is measured usually below 0.045% and for short membranes, the strain is above this mean value.

**Conclusion**

In summary, we have achieved nanostructures in monolayer, with monodomain $MoS_2$ suspended on $SiO_2$ nanopillars. Using Raman spectroscopy, optomechanical and photoluminescence measurements, we have not only investigated the different contributions from the strain, doping, temperature and layer number but also clearly separated each component. We were able to achieve a clean array of periodically strained $MoS_2$ and measured an assembly of mechanical resonators in this 2D material. Under these conditions, we have shown the resonator frequency to be homogeneous along the sample for a specific geometry. In order to confirm our protocols and emphasize the doping, strain and temperature effects, we tested samples with 2R≈a and 2R<<a for different laser beam powers. In the limit of 2R<<a, where the $MoS_2$ forms a tent structure, we found a very strong downshift in the optical band gap from 1.89 to 1.45 eV, which originated from the strain. This is partially due to a shift of the direct band gap transition, but also due to the appearance of an indirect band gap. This effect has been optically observed, thanks to the specific geometry and configuration of the devices. For fully suspended 2R<a membranes, we were able to extract the doping difference Δn and the thermal properties. From the same measurements, taking advantage of the proportionality between doping and strain effect on Raman peaks, we estimated the strain increase at the pillar apex to be about 0.5%. All these results on $MoS_2$ nanoengineering can contribute information to the optoelectronic field related to the creation of local quantum emitters with TMDs on pillar arrays, [10,11,23] where strain, doping, and temperature are essential parameters. Further studies will explore the mechanics and the possibilities presented by different types of pillar distribution, shedding light on the importance of correlating the geometry of the sample with the strain distribution or the vibration properties. Efficient control of the strain or spring constant with nanostructuring, together with other properties such as doping or temperature, can potentially enhance the properties of force or mass sensitivity within nanomechanical resonators. [3,47]

Finally, we report a method to extract the thermal conductivity of our sample. Recent results [4–6] indicate that thermal transport is quite challenging in 2D materials and the record of thermal conductivity [7–9] only represents a part of the phonon mechanism in 2D materials; a major and fundamental aspect is the length dependence of the thermal conductivity in these materials on the nanoscale. Controlling thermal transport or local vibrations by tailoring the geometry of the system in suspended samples is at the heart of phononics, which today relies mostly on silicon thin film technologies. The extension of this concept to thermal transport in suspended and nanostructured 2D materials will permit the combination of thermal transport engineering with the highest thermal conductive materials. [8]

**Acknowledgments:** This work was supported by ANR H2DH grants, by the European Union's Horizon 2020 research and innovation program under grant agreement No 732894 (FET Proactive HOT) and by the French Renatech network.

**Supporting Information:** The Supporting Information is available free of charge on the ACS Publications website. Additional information includes a reference table for $MoS_2$ Raman peaks and photoluminescence dependence with strain, temperature, and doping, some additional devices of tent structures and large membranes It details also the procedure for simulations, an additional calibration for the temperature effect, some doping effect on Raman measurements. We have included the mechanical model for our 2D drums

**Conflicting financial interests:** The authors declare no conflicting financial interests.

**Supplementary information for:**

**Intrinsic properties of suspended MoS$_2$ on SiO$_2$/Si pillar arrays for nanomechanics and optics.**

Julien Chaste[1*], Amine Missaoui[1], Si Huang[1], Hugo Henck[1], Zeineb Ben Aziza[1], Laurence Ferlazzo[1], Adrian Balan[2], Alan. T. Charlie Johnson Jr.[2], Rémy Braive[1,3], Abdelkarim Ouerghi[1]

[1] Centre de Nanosciences et de Nanotechnologies, CNRS, Univ. Paris-Sud, Universite Paris-Saclay, C2N – Marcoussis

[2] Department of Physics and Astronomy, University of Pennsylvania, 209S 33rd Street, Philadelphia, Pennsylvania 19104 6396, United States

[3] Université Paris Diderot, Sorbonne Paris Cité, 75207 Paris Cedex 13, France

* Corresponding author: julien.chaste@c2n.upsaclay.fr

**S1: Examples of nanostructuring with nanopillars**

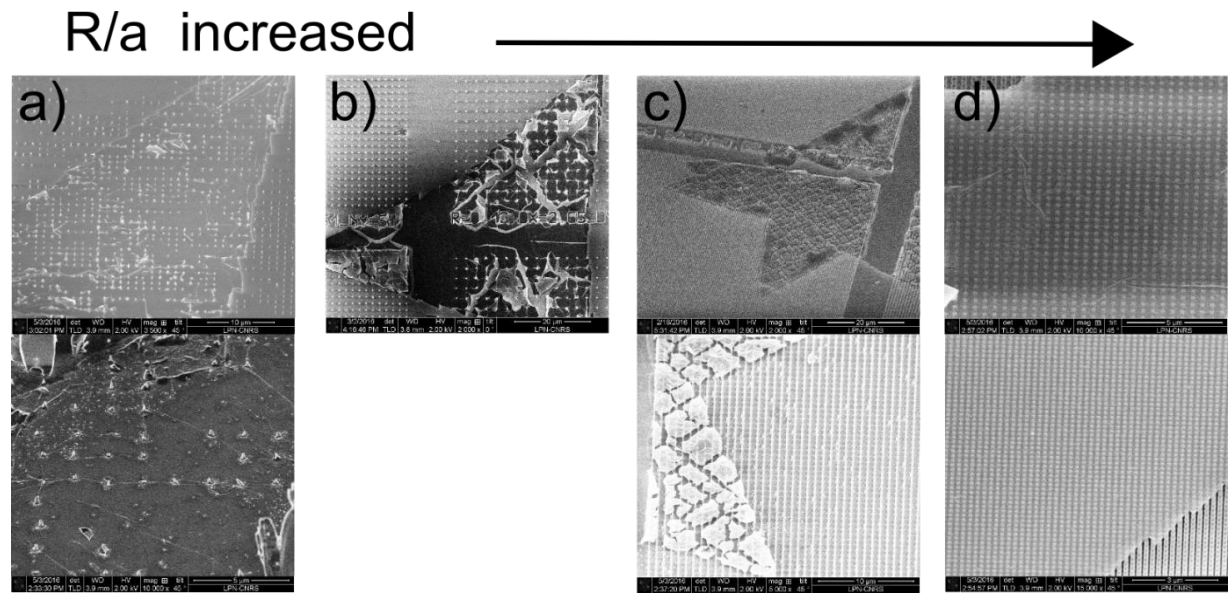

**Figure S1a:** Electron beam images of MoS$_2$ on nanopillar arrays with different Radii, R and period a. From a) to d) the R/a ratio increases and, consequently, the mechanical stability of the suspended structure against cracking and ripples. In a) we can see the same type of ripple organization as in the reference [1]. The shape and structure of membrane cracks and ripples are correlated with the pillar structure.

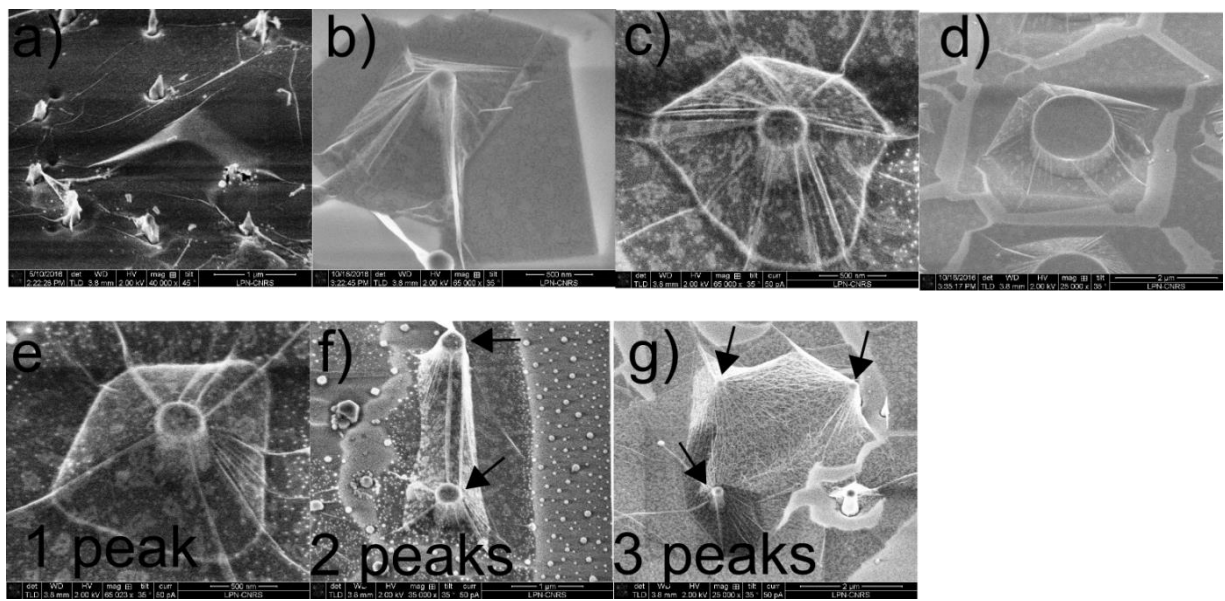

**Figure S1b: Tent structures with tall pillar height.** In the regime defined in Figure S1a-a, we see a large amount of "tent" formed by MoS2 covering pillars with different pillar radii and heights. In a) to d), the radius increases. It is possible to build tent structures with heights of 600-820nm in comparison with ref [1–3] the height does not go above 200 nm. In e) to g), we see different tent structures with two or three pillars.

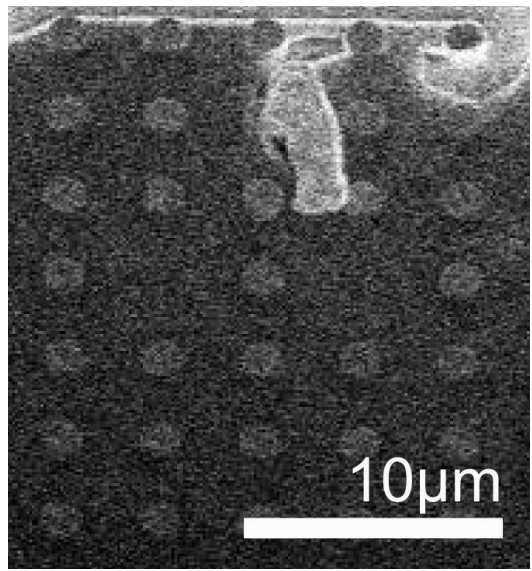

**Figure S1c: Very large suspended structures.** When the transfer process is optimized, our method creates arrays of $MoS_2$ square membranes with 6.5μm sides.

**S2: MoS2, Raman, and photoluminescence**

The table shows citations for the Raman peaks $E^1_{2g}$ and $A_{1g}$ and the photoluminescence peaks A, B, and I) as a function of doping, strain, and temperature for different layer numbers. The peak shift depends on the substrate interaction, the nature of stress applied to the sample, the number of layers or the temperature.

**Raman peak and photoluminescence peak behavior in relation to strain, doping and temperature**

| N-layer | Peaks | Strain (cm-1/%)    strain (meV/%) | Doping ($/10^{13}cm^{-2}$) | Temperature cm$^{-1}$/100K |
|---|---|---|---|---|
| 1 | $E_{2g}$ | **-5.2(bi,sus)**[4],-4.6(**bi**)[5],-2.1 (uni) [6],-4.5/-1(uni)[7],-1.3(uni)[8],-2.5/-0.8[9] | -0.33[10] | -1.3[11], |
|   | $A_{1g}$ | **-1.7(bi,sus)**[4],-1(**bi**)[5], -0.4(uni) [6],-0.4(uni)[8] | -2.2[10] | -1.6[11] |
|   | A | **-99(bi,sus)**[4],95(**bi**)[5],-45(uni)[7],44(th)[12],-25[8],-64[13],-48[14] | -70(th)[15]*, -66[16]* | |
| 2 | $E_{2g}$ | **-5.2(th,bi,sus)**[17],-4.6/-1(uni)[7+] | | |
|   | $A_{1g}$ | **-2.2(th,bi,sus)** [17] | | |
|   | A | **-120(th,bi,sus)** [17], **-91(bi,sus)**[4],-53(uni)[7],-48[13],-46[14] | | |
|   | I | **-290(th,bi,sus)** [17], **-144(bi,sus)**[4],-120(uni)[7], -77[13],-86[14] | | |
| N | $E_{2g}$ | **1.7(uni,sus)**[18],-1.7(uni)[6],-3.7(uni)[8] | | -1.5[11],-1.32[19] |
|   | $A_{1g}$ | -0.4(uni) [6],-0.7(uni)[8] | | -1.3[11],-1.23[19] |
|   | A | **-73(bi,sus)**[4],--60[8] | | |
|   | I | **-110(bi,sus)**[4] | | |

*=indirect deduction from graph and text, +=  E+ et E- measurements after splitting,
Sus=suspended, Bi=biaxial, uni=uniaxial

**S3: MoS2 on pillars: in relation to doping**

In our samples, part of the MoS$_2$ was suspended and part was on the SiO$_2$ surface, at the SiO$_2$ pillar apex, for example. The first well-known characteristic of a system such as a 2D material is the strong doping dependence and interaction with the substrate, especially in the case of MoS$_2$. [20] If the material was suspended with a doping $n_s$ or on a SiO$_2$ substrate with a doping $n_n$, the Fermi level was seen to shift up to 40meV. [17]

The case of doping is easier to simulate than strain and heating because this depends only on $n_n$, $n_s$ and D: the doping in the non-suspended and suspended parts and the laser spot size, respectively. In addition, both $n_n$ and $n_s$ must be quite similar from one sample to the other. In order to define $n_s$ and $n_n$, it is possible to measure a reference to a large suspended sample and a large non-suspended part, as in ref [17] for bilayer MoS$_2$. In our case, we needed a calibration *in situ* due to the small deviations in the reference peak position because of doping ($\leq$1cm$^{-1}$). Our problem was that the doping effect was reduced to an effective doping because the pillar dimensions were smaller than the spot dimensions. The effective doping had an intermediate value between $n_n$ and $n_s$. The effect of doping on Raman peak concerned mainly the A$_{1g}$ peak.

In periodic patterns such as in Figure 3, with R=0.215μm and a=0.5μm, we could measure modulation of the A$_{1g}$ peak position. We extracted the effective peak intensity modulation, defined as the ratio $I_{max}/I_{min}$ and dP, which is the effective difference in peak position at maximum and minimum intensities. This modulation came from doping, as presented by the Lee *et al* diagram [21] for the two peak positions (Figure S3c). The photoluminescence intensity of the A peak also shows a strong modulation without energy shift of the peaks or intensity variation of the A- and B peaks (Figure 3), being a signature of the doping variation in the MoS$_2$ along the structure.

For doping simulations, we assumed an initial Lorentzian peak for the suspended MoS$_2$ with a typical width W$_S$=5.5cm$^{-1}$, and position P$_S$=405cm$^{-1}$ for the A$_{1g}$ peak (height H$_S$ set to 1 by default for the undoped suspended situation). We fixed the Raman peak dependence as a function of ref [10], taking df=4.(n$_n$-n$_s$)/1.8, for H=1+n$_{n,s}$/1.8 , W=5.5+n$_{n,s}$.(12-5.5) (df in cm$^{-1}$, n$_n$ in 10$^{13}$cm$^{-2}$ H in u.a., and W in cm$^{-1}$). The Gaussian spot size was around 300-400nm in most of our data but remaining a free parameter in our simulations. We calibrated this at the edge of the MoS$_2$ (Figure S3b). With the measurement of $I_{max}/I_{min}$ and df, it was possible to define D, n$_n$ and n$_n$. In the case of Figure 3, we found, after convolution of local mapping for the measured photon intensity with the Gaussian spot size, $I_{max}/I_{min}$ =1.25 and dP=0.28cm$^{-1}$. This gave a real peak position difference df of 1.1cm$^{-1}$, a doping difference $\Delta n = n_s - n_n = 5.10^{12} cm^{-1}$, a real width $W_n = 7.3 cm^{-1}$, an intensity $H_n = 1.72$ and a Gaussian spot size of D=300nm.

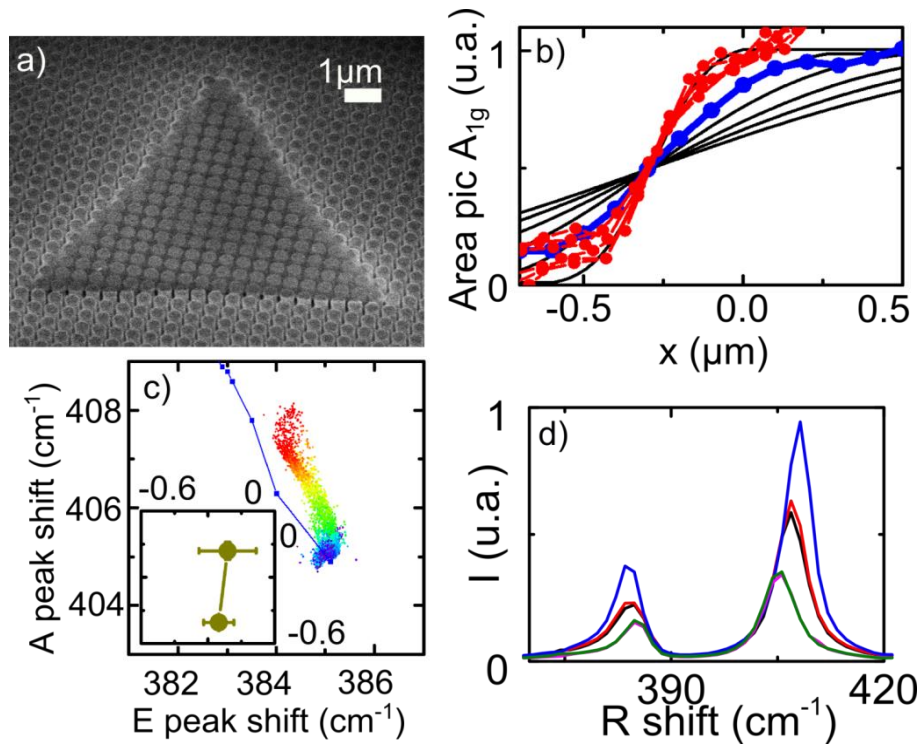

**Figure S3a: MoS$_2$ flake on top of pillars with 2R~a: layer number variation and Gaussian spot size calibration a)** e-beam image of a MoS$_2$ flake with different layer numbers. b) The Gaussian spot size calibration with A$_{1g}$ peak intensity in blue and red taken at the red and blue bar positions in Figure S3b. In black, the fit of Lorentzian area, along with a step in intensity, correlated with a 2D Gaussian spot size, normalized at 1 (Diameters of 200, 400, 600, 800, 1000, 1200nm). The spot size is between 300 and 400nm. c) Diagram of the respective peak positions of MoS$_2$, for the whole points present in Raman mapping. This corresponds closely to the variation in the number of layers (blue line). In the inset, we drew the same diagram along the peaks for suspended and non-suspended parts in Figure 3d after background removal and averaging. A doping variation was identified. In d, the different spectra for 1 layer, 2 layers, and N layers.

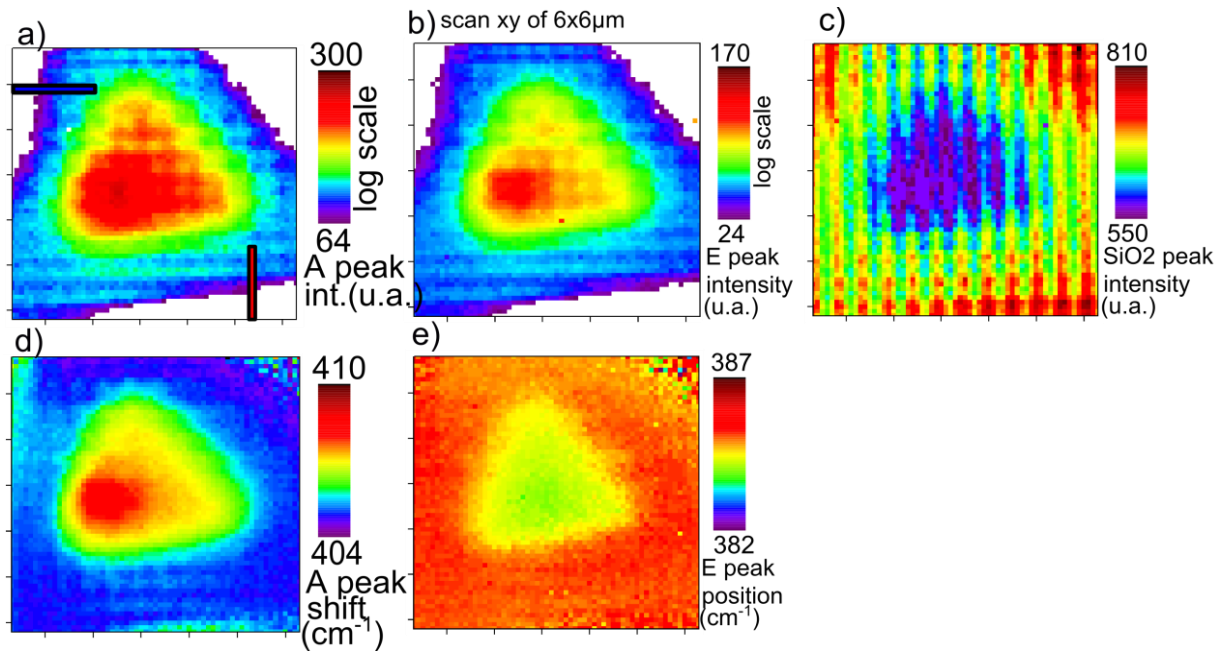

**Figure S3b: MoS2 flake on top of pillars with 2R~a: additional data for Figure S3b.** Raman mapping of $A_{1g}$ and $E^1_{2g}$ data for intensity (a, b) and peak shift (d, e) In (d) and (e) we have raw data on the left and background subtracted data on the right. In (c), the SiO$_2$ peak intensity defines the peak position. A modulation of the shift with pillars is only observed for the $A_{1g}$ peak; this indicates a doping modulation.

## General relationship between material properties and Raman peak shape when a nanostructure and a high gradient are present

In order to determine properly the real local peak position and intensity in the suspended and non-suspended parts from the convoluted measured data, we used the following two methods: a simple and intuitive method with a strong approximation and an exact methodology. The approximated method is empirically valid for small or gentle variations, i.e. when the peak shift is less than 1cm$^{-1}$ and can diverge for greater variations.

First, we define an xy map for the different property variations. The easier method is simply to convolute this mapping with a 2D Gaussian distribution at each point and to transform it into frequency shift or peak intensity. To be exact, we measured all the collected Raman photons, with a Gaussian distribution around the spot center. This meant we had to convolute the Raman intensity at each Raman frequency (cm$^{-1}$) after the peak intensity at each point had become modified by the local strain, doping or temperature. This is much more precise and general, but it is not intuitive, because the peak can have a non-standard shape if the gradient shift is too high and the peak variation mapping is not necessarily linear with the doping, strain or temperature mapping.

For the sample inFigure 1c, R=0.13µm, and a=1.21µm. We estimated, in a first step, the doping to be $\Delta n = n_s - n_n = 5.10^{12} cm^{-1}$. On a second step, we tested different doping differences $\Delta n$ and found a result which can converge only for doping values around $=5.10^{12} cm^{-1}$ (see Figure S4d). This confirms the previous calibration.

The value of $\Delta n = n_s - n_n = 5.10^{12} cm^{-1}$ was quite unexpected because the suspended part of the $MoS_2$ is usually less doped due to less interaction with the substrate and without an external doping source. Our result means our $MoS_2$ is naturally doped and the doping is reduced by the $SiO_2$ interaction. To explain this, we have to consider the specific case of our $MoS_2$. In Figure S3d, we can see $MoS_2$ flakes that are not really triangular. This is a result of sulfur vacancy in the $MoS_2$, as described in ref.[22] We have previously measured the natural doping of the $MoS_2$ appearing in our devices , [23,24] which is not negligible. In addition to this, we carried out electrical measurements on a suspended $MoS_2$ flake between 2 electrical contacts. As shown in Figure S3d, there is strong photogating appearing in the suspended $MoS_2$, even with a low power laser. The efficiency of this photogating, compared to non-suspended devices with the same $MoS_2$ [24] is the proof that our devices can be more doped in the suspended than in the undoped parts, with the laser illumination lower than in Raman spectroscopy, where the laser power is usually around 50µW.

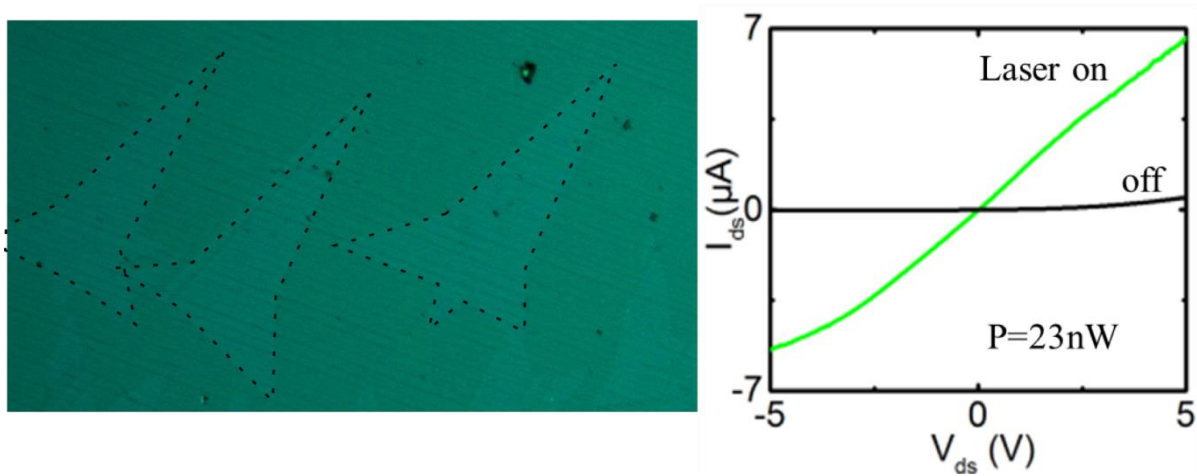

**Figure S3d: typical $MoS_2$ with sulfur vacancies and doping in suspended samples.** An optical image of typical $MoS_2$ flakes with convex edges. This is a signature of sulfur vacancies in our $MoS_2$ and can explain a natural doping by dopants localized in $MoS_2$ at the sulfur

vacancies. An I(V) curve of a typical suspended MoS$_2$ flake with and without illumination at 532nm. We observed a dark state in the OFF position and an Ohmic conductance in the ON position, even with only 23nW. This means we activated strong photogating in our devices, which can explain the doping difference $\Delta n = 5.10^{12}$cm$^{-2}$ appearing in our Raman measurements.

**S4: MoS2 on pillars: in relation to heating**

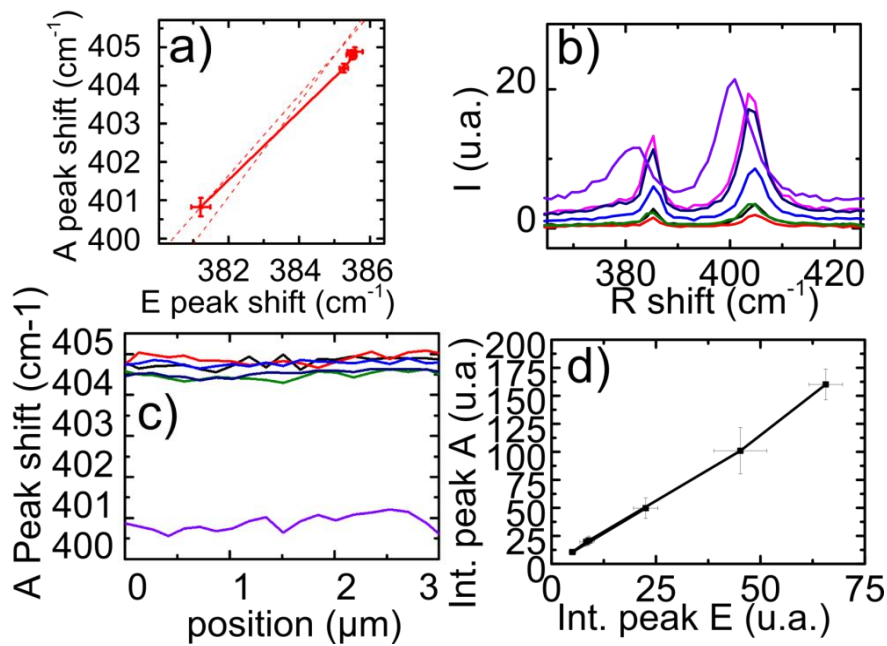

**Figure S4a: Laser heating effect on suspended monolayer MoS$_2$.** b) Raman spectral measurements on a MoS$_2$ flake suspended between pillars in the limit 2R<a, and with a geometry very similar to the sample in Figure 1c (P=5, 10, 50, 100, 500µW). The data was taken at the center of the suspended area, far away from the pillars. c) The $A_{1g}$ Peak position along the suspended $MoS_2$ as a function of laser power. This is quite homogenous. a) A Lee at al. diagram of the respective peak positions for the average points in c and for every laser power intensity. d) The same diagram with the peak intensity. Both confirm a shift due to heat in the sample, with a slope close to 1.1 for the position.

In order to determine the heating effect on our samples, especially the ratio $\Delta A1_g /\Delta E^1_{2g}$ we used a very simple methodology. We used a sample of suspended monolayer $MoS_2$ on $SiO_2$ pillars similar to the sample in Figure 1b. We placed the laser between pillars and we measured the Raman signal as a function of the laser power, especially the peak position and the ratio $\Delta A1_g /\Delta E^1_{2g}$. We repeated the same measurement along a suspended membrane (each point being away from the pillars) to determine the small standard deviation of our measurement. We did observe a broadening of the Raman peak at high powers and a linear variation between the two Raman peak intensities, as expected from the heating variation of the $MoS_2$ membrane. This method involves some approximations, but the result is completely in accordance with the literature and a $\Delta A1_g /\Delta E^1_{2g}$ ratio of around 1.1.

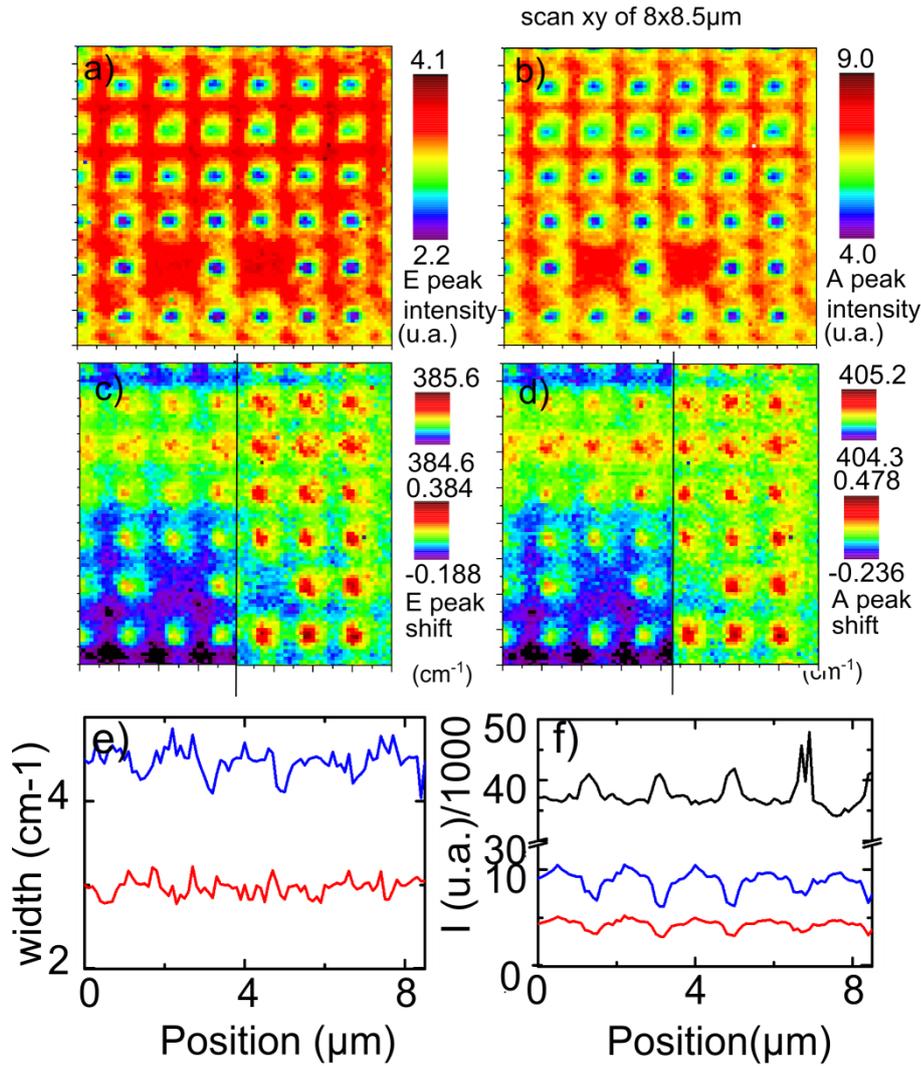

**Figure S4b: Figure 1c sample: additional data.** A) to d) Raman mapping characterization with peak intensity and position modulation along the sample. We observed a correlation between pillar position and peak modulation. c) and d) we removed, on the right part, a background deviation due to 15 hours of measurements which does not appear in a fast measurement at the exact same position as in e, f, and Figure 3a. e) and f) the peak width and intensity of the sample. We observed a modulation of the $A_{1g}$ peak width with peak appearance which is certainly due to the doping signature and $MoS_2$ modulation along the pillars. As a first approximation, in the linear variation, we do not expect the heat and strain to affect the width of the $A_{1g}$ peak alone.

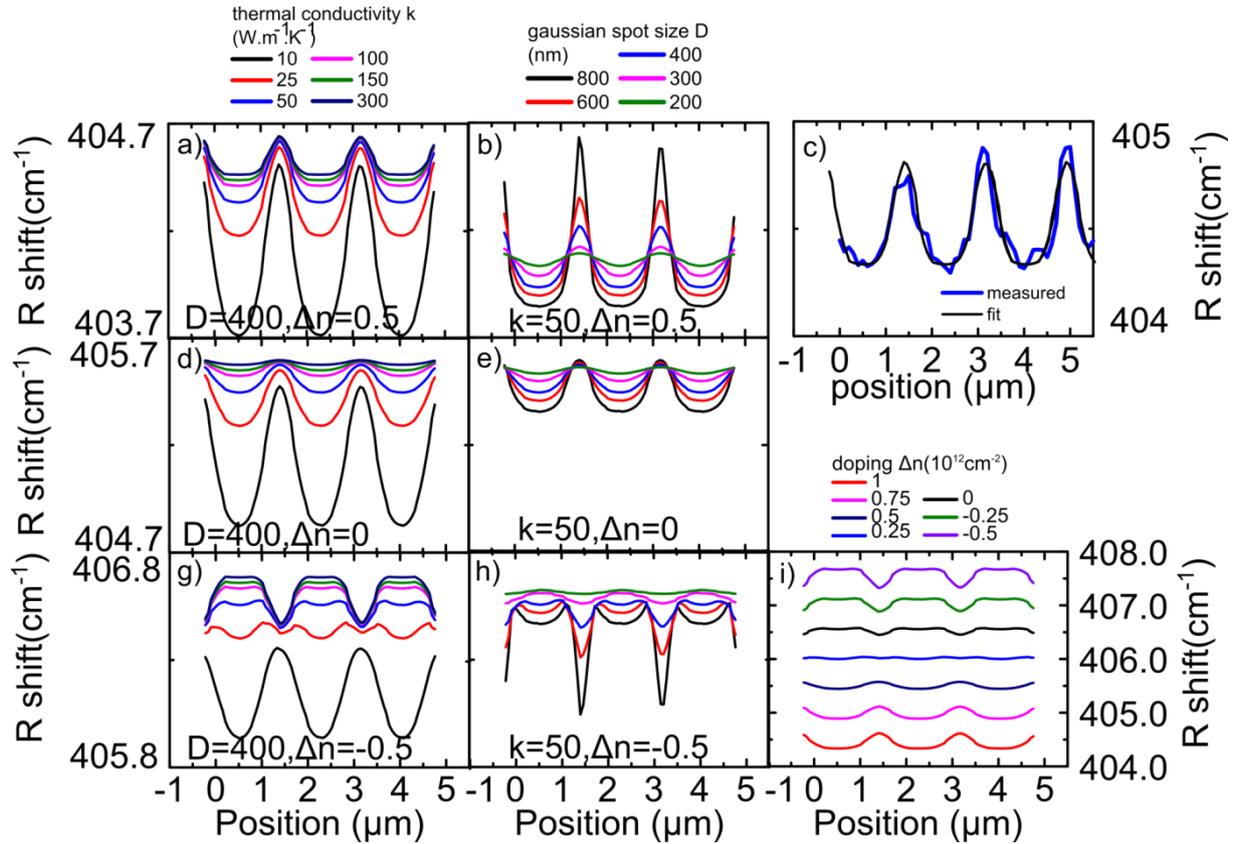

**Figure S4c: Figure 1c sample: additional simulation with doping and heating.** We used by default the best parameters for the best fit; the doping difference $\Delta n=0.5 \cdot 10^{12}$ cm$^{-2}$, the thermal conductivity $k=50$ W.m$^{-1}$.K$^{-1}$, and the Gaussian spot size D=400nm. The data and the best fit are shown in c). In the other panel, we have modified one of the parameters; in a), d), g), thermal conductivity, in b), e) and h), Gaussian spot size and in i), doping difference. In a) and b), d) and e), g), and h), we fixed doping values of -0.5, 0 and 0.5 ($\cdot 10^{12}$cm$^{-2}$), respectively. It is interesting to note that, in d) and e), we have the result without taking into account the doping effect and this is not sufficient to explain the specific shape of our data.

For the COMSOL simulation, presented in Figures 3a and b, of the thermal heat transport in MoS$_2$, we used:

- density of 2163e-6kg/m3,
- heat capacity of 138 J/K/kg,
- absorption coefficient of 6% per layer.
- number of layers=1
- layer thickness of 1.12nm
- laser power of 50μW with a 2D Gaussian distribution.

## S5: MoS₂ on pillars: the optical cavity created between MoS₂ and SiO₂

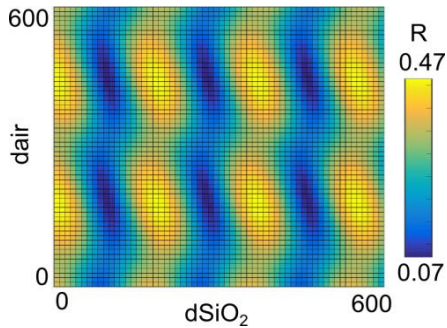

**Figure S5: Reflection of the laser modulated by the air cavity distance and SiO₂ thickness.** We simulated the reflection of our system. We defined $d_{air}$ to be the distance between the MoS₂ monolayer and the SiO₂/Si substrate and $d_{SiO2}$ the thickness of SiO₂. Calculated with the MATLAB Multidiel2 function.

For the Raman and photoluminescence measurements, the reflectance amplitude is modulated by the cavity created between the MoS₂ and the substrate. The space between MoS₂ and the substrate, $d_{air}$, can be seen as an optical cavity, a very poor one, but an optical cavity all the same and that influences strongly the laser reflection at 532nm. The signal will oscillate between minimum and maximum values, with respect to some periods in $d_{air}$. In any case, because the cavity is quite poor, it will not enhance the signal on the suspended part drastically for any $d_{air}$ values and does not, by itself, explain the emergence of a strong signal between 1.3 and 1.6eV in Figure 2.

These modulations are important for samples where this distance $d_{air}$ varies a lot, as for the "tent structures". $d_{air}$ goes from 800nm to 0 in less than 1μm. In order to determine the laser reflection with accuracy at any position, we simulated the reflectance of a laser at 532nm on a sample made from one monolayer of MoS₂, an air thickness $d_{air}$, a SiO₂ layer of thickness $d_{SiO2}$ and a Si substrate. We used a model described in ref [25] to plot the reflectance as a function of $d_{air}$ and $d_{SiO2}$. The reflectance is periodic, with both parameters having high amplitudes of oscillation from 47% to 7%.

Nevertheless, in Figure 2, $d_{air}$ can change from 600-800nm to 0 in almost 0.5μm laterally, along x. This is almost the laser spot size. As mentioned previously, the laser is of a

Gaussian spot size with a diameter around 400nm at best. this means the cavity effect and the fast modulation (2 periods in Figure S5) are completely smoothed out over the sample and we measured an average value which corresponded to a value without any cavity modulation in these "tent "structures.For other samples, this effect can be important in case of bending or any curvature of the MoS$_2$.

## S6: Optomechanics on a network of mechanical resonators with MOS$_2$ membranes deposited on pillars

Within our MoS$_2$ samples, mechanical resonators at some positions are basically defined by the pillars near this position. This means that they are defined by 4 pillars for the small resonator and 8 for the large resonators (membranes where a pillar is missing in the middle). It is possible to simulate the motion and resonance frequencies with finite element analyses, but we used a simpler analytical model by approximating our resonator to a circular drum resonator.

We used the Euler Bernoulli model of a membrane to solve the resonance frequency f$_0$, as clearly described in ref [26] for a drum;

$$f_0 = \frac{kd}{4\pi} \cdot \sqrt{\frac{16D}{\rho d^4} \cdot \left[\left(\frac{kd}{2}\right)^2 + \frac{\gamma d^2}{4D}\right]}$$

With $D = \frac{Et^3}{12.(1-\nu^2)}$, E=0.33 Tpa, the Young modulus of the material, $\nu = 0.125$ the Poisson ratio, d the diameter of the membrane, kd/2~2.4048 a coefficient related to geometry, $\gamma$ the strain, $\rho = 3073$ kg.m$^{-2}$ the 2D mass density, and t=6,15.10$^{-10}$m, the monolayer thickness.

The result is described in Figure 5b and fits our data well for a strain between 0.035% and 0.055% (grey area). This means our sample resonances are dominated by the strain, with a spring constant around 0.6 N/m.

## S7: MoS$_2$ on substrate: Raman calibration of heating and doping

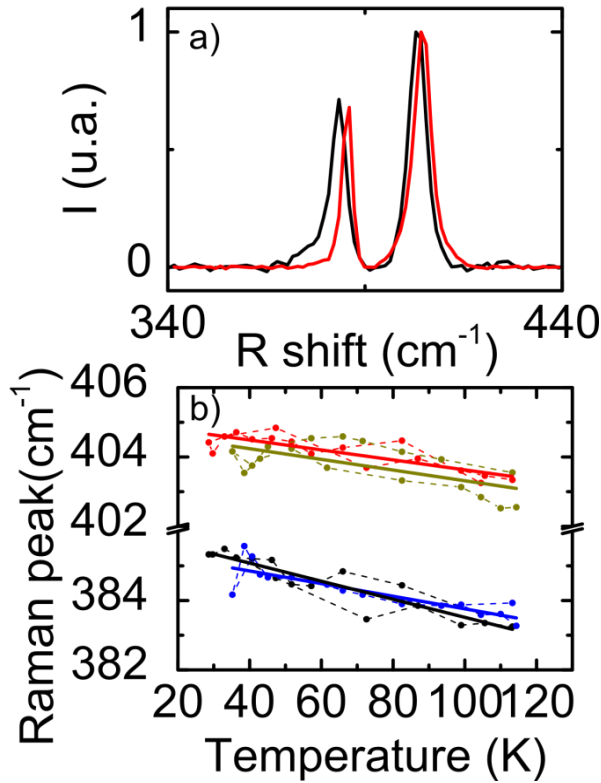

**Figure S7a: Calibration of the Raman spectrum in function of the temperature.** We have measured the spectrum of different flake of monolayer MoS$_2$ in function of the temperature directly on the growth substrate of SiO$_2$/Si. a) Raman spectrum for 30°C (red) and 113°C (black). b) Raman peak position of the A$^1_g$ and E$^1_{2g}$ peak position in function of the Temperature with the thermistor at two different places on the hot plate and with back and forth measurements in temperature.

To confirm the quantitative analyze of the thermal conductivity, we did a temperature-Dependent Raman Studies for both E$^1_{2g}$ and A$^1_g$ peak dependence. In figure S7b, we measured a $\Delta A^1_g/\Delta E^1_{2g}$ ratio of -0.014cm$^{-1}$/K and -0.015cm$^{-1}$/K for the A$^1_g$ peak and -0.018cm$^{-1}$/K and -0.026cm$^{-1}$/K for the E$^1_{2g}$ peak. For the A$^1_g$ peak these value are both comparable with previous results in the literature (-0.016cm$^{-1}$/K). [11,19] For the E$^1_{2g}$ peak, we can see a deviation from previous results measured at -0.013cm$^{-1}$/K and with our measurements of heating in function of the laser power (see part S4). This small difference can be explain by an additional strain in our sample considering the different flake of MoS$_2$ used for both measurements; We attribute this strain to the thermal expansion of both layer-substrate system and from different mechanical coupling with the substrate: this is different if the flake stick to the substrate or slide on the substrate (thermal expansion of monolayer TMDs is $1.10^{-5}K^{-1}$ and of Si is $2.10^{-6}K^{-1}$; for 100°C, it can eventually modify the length of the flake by 1% depending of the coupling nature with the substrate). Our measurements were done at ambient conditions with a TH10K thermistor and HT24S ceramic heater from Thorlab and decouple with Kapton layer and a Teflon plate (Figure S7c). We took care to work at the exact same position between each point and to stabilize the temperature of the system with sufficiently long time scale. To confirm our temperature measurements, we have done two measurements, on two different $MoS_2$ flakes. Each time, we have placed the thermal coupler at two different places on the heater (brown and red curve in S7b for example) and we have not seen any notable difference in the $A^1_g$ peak dependence. We have measured with P=100μW in order to be sure of the absence of heating from the laser itself.

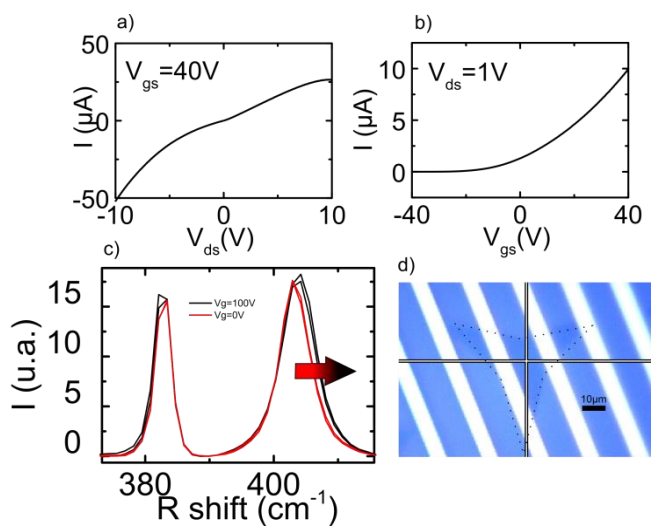

**Figure S7b: Calibration of the Raman spectrum in function of the doping for a $MoS_2$ flake on $SiO_2$/Si substrate.** a) and b) we have measured the I(V) curve with a probe station. c) An image of the set-up under Raman spectrum. d) An image of the $MoS_2$ flake under test connected to interdigitated metallic pads. The crossbar indicates the laser position. e) Raman spectrum of $MoS_2$ with different gate voltage of 0 and 100V at ambient conditions with a laser power of 100μW. We observe a shift of the peak and a broadening with doping. We proceeded twice to the same measurement and we have seen the good reproducibility of our results.

We have also done some Raman spectrum measurement of our $MoS_2$ under different electrical polarization in order to calibrate the Raman doping dependence and confirm our statement on this point. We have contacted, by e-beam lithography, a $MoS_2$ flake on a substrate with interdigitated gold pads with the substrate acting as a bottom back gate. The dielectric thickness is 300nm and the gate capacitance $C_g$ is 8.9nF.cm$^{-2}$. We have measured the I-V characteristic with a probe station and an Agilent 4155C at ambient conditions under white light illumination. The MoS$_2$ is naturally n-doped with the Fermi level closed to the semiconducting band gap. We have proceeded with Raman measurements at different gate voltage (with a Keithley 228A). We measure an upshift of the A$^1_g$ peak of 0.7cm$^{-1}$ with a broadening of the peak between $V_g$=0V and $V_g$=100V. On contrary the E$^1_{2g}$ peak is not affected by doping. All these behaviors are expected for doping variation if we consider the value measured in the literature of 2.2cm$^{-1}$ for A$^1_g$ peak and a doping shift of 10$^{13}$cm$^{-2}$. Our shift corresponds to a doping change of 3.10$^{-12}$ cm$^{-2}$. If we consider the relation between charge density and the capacitance n.e=$C_g$.$\Delta V_g$ with e the electron charge= 1.6.10$^{-19}$C and $\Delta V_g$ the gate shift, we can estimate the doping change to be around 5.10$^{12}$cm$^{-2}$; not very far from the measured value. We confirm our data analyses trough this calibration.

**SOM References**